\documentclass[10pt,a4paper]{article}
\usepackage{times}
\usepackage{graphics}
\usepackage{graphicx}
\usepackage{float}
\restylefloat{figure}

\begin{document}
\large
\date{ }

\begin{center}
{\Large  Experimental search for the bound state singlet deuteron in the
radiative $n-p$ capture.}

\vskip 0.7cm

T. Belgya$^{a}$, S.B. Borzakov$^{b,d}$, M. Jentschel$^{c}$, B. Maroti$^{a}$,
Yu.N. Pokotilovski$^{b}$\footnote{corresponding author,
e-mail:pokot@nf.jinr.ru}, L. Szentmiklosi$^{a}$

\vskip 0.3cm

        $^{a}$ Centre for Energy Research, Hungarian Academy of Sciences\\
             H-1525 Budapest 114., P.O. Box 49., Hungary

\vskip 0.3cm

             $^{b}$ Joint Institute for Nuclear Research\\
              141980 Dubna, Moscow region, Russia\\

\vskip 0.3cm

              $^{c}$ Institut Laue-Langevin,\\
           BP 156, 38042 Cedex 9  Grenoble, France

\vskip 0.3cm

             $^{d}$ University "Dubna",\\
              141980 Dubna, Moscow region, Russia\\

\vskip 0.7cm

{\bf Abstract\\}

\begin{minipage}{130mm}

\vskip 0.4cm

 We performed an experimental search for the bound state singlet deuteron
predicted in some microscopic calculations.
 The predicted energy of this metastable level is in vicinity of the deuteron
disintegration threshold.
 This state should manifest itself in two-photon transition following thermal
neutron capture by protons.
 The experiment consists in the search for the second gamma-ray in the cascade
through a high statistics measurement of $\gamma$-ray spectra after cold
neutron capture by hydrogen nuclei.
 The upper limit $2\mu$b (2$\sigma$ level) is obtained for the cross section of
the singlet deuteron production with the bound energy in the range  10-125 keV.

\end{minipage}
\end{center}

\vskip 0.3cm


\vskip 0.2cm


\vskip 0.6cm

\section{Introduction}
 Singlet deuteron (S=0, T=1) is usually considered as not bound, but as a
virtually bound ("antibound", "quasibound") state with binding energy B$<$0
indicating unstable configuration \cite{Ma}.
 A number of experiments was devoted to search for this state
(denoted as d$^{*}$) in different nuclear reactions.

 Cohen et al.\cite{Coh,Coh1} observed singlet deuteron in the reaction
$^{9}$Be(p,d$^{*}$)$^{8}$Be at the energy of incident protons 12 MeV.
 Bohne et al. observed analogous process in the reaction
$^{3}$He($^{10}$B,d$^{*}$)$^{11}$C \cite{Boh}.
 Gaiser et al.\cite{Gai} investigated the reaction $^{4}$He(d,p$\alpha$)n at
the energy of bombarding deuterons 7 MeV.
 Their data gave clear evidence for the production of the unbound singlet
deuteron d$^{*}$.

 Bochkarev et al.\cite{Boch} investigated decays of excited 2$^{+}$ states of
the $^{6}$He, $^{6}$Li, $^{6}$Be nuclei.
 From the energy and momentum conservation the narrow peaks in the $\alpha$-
spectra were considered as indications of the two-particle decays:
an $\alpha$-particle and the singlet deuteron in the case of $^{6}$Li and
an $\alpha$-particle and the dineutron in the case $^{6}$He.
 Interpretation of the experimental spectra in terms of the two-nucleon
final state interaction have lead to an abnormally large nucleon-nucleon
scattering length $\sim$50-100 fm.

 Generally the problem of existence of the singlet deuteron is closely
connected to the old problem of existence of the dineutron and more generally
of the neutral nuclei.

 Experimental search for the dineutron was the subject of a number of
experiments \cite{din}.
 In some of them there were indications of observation of the dineutron,
the tetraneutron \cite{tetr}, and even multineutrons with a number of neutrons
n$\ge$6 \cite{mult}.

\section{Theoretical indications}
 As is known it is impossible to conclude from the np-scattering experiments
whether the state is virtual or bound.
 We can cite from the respected monograph: "Only if there is no bound state
capable to account for the low-energy cross section one is entitled to give
definite statements about the existence of antibound states" \cite{Alfaro}.

 Over the years the claims have appeared that the binding energy of the $np$
pair in the singlet state may be positive, the singlet deuteron is stable in
respect to decay to the neutron and proton.

 Maltman and Isgur \cite{Malt} described $np$ system as the six quark state,
they have obtained 400$\pm$400 keV binding energy for
the singlet state and 2.9 MeV for the triplet state.

 Ivanov et al. \cite{Iva} considered the deuteron as the Cooper np pair in
the field-theoretical approach developed within Nambu-Jona-Lasinio model
of light nuclei.
 They computed a binding energy of
$\epsilon_{S}$=79$\pm$12 keV for the deuteron singlet state modeled as Cooper
pair in the $^{1}$S$_{0}$ state and calculated the S-wave np scattering length
in terms of the binding energy.
 The calculations agree well with the energy of the virtual level
$\epsilon_{S}$=74 keV, defined from the experimental S-wave scattering length.

 Hackenburg \cite{Hack} employed the intermediate off-shell
singlet and triplet deuterons treated as dressed dibaryons in his calculations.
 In simple extension of the effective range theory he predicted existence of
the singlet deuteron bound state.
 The binding energy of this singlet level was predicted to be E$_{S}$=66 keV.
 He showed that the radiative capture leads to the possibility
of observation of the metastable singlet level in the resonance scattering of
gamma quanta by deuterons and in two-photon radiative capture with the expected
cross section for the latter 27$\mu$b, more than four orders of magnitude less
than the main np radiative capture channel.

 Calculations of Yamazaki et al \cite{Yam} in the quenched lattice quantum
chromodynamics lead to conclusion that not only the triplet but also the singlet
deuteron state should be bound.

 It is also possible to use the idea of negative resonance instead of the
virtual level \cite{Ma,Borz,Borz1} as phenomenological model for the $np$
scattering.
 Shapiro et al. \cite{Shap} used this approach to describe the radiative
capture reaction n($^{3}$He, $^{3}$H)p.
 It was confirmed in the proton-tritium scattering experiment of Gibbons et al. \cite{pT}.

 In Ref. \cite{Borz} it was assumed that the imaginary part of the scattering
amplitude corresponds to the radiative resonance width.
 This model can also treat the scattering and the radiative capture.
 According to this model the resonance width is of the order 10 eV, and the
bound state singlet deuteron may be observed in the resonance gamma-ray
scattering by deuterons or in the radiative $^{1}H(n,\gamma$) capture.

\section{Experiment}
 To search for the bound deuteron singlet state gamma ray pulse height spectra
were recorded from the radiative neutron capture by protons.
 Existence of the bound state singlet deuteron could be evidenced by a two-step
gamma-ray transition $^{3}$S$_{1}$ (continuum) $\rightarrow^{1}$S$_{0}$
(metastable)$\rightarrow^{3}$S$_{1}$ (ground state) in addition to the direct
one $^{1}$S$_{0}\rightarrow^{3}$S$_{1}$  with energy 2223 keV .
 The main interest was concentrated in the energy range $\sim$100 keV below
the gamma ray with E=2223 keV.

 Preliminary experiments performed at the Dubna pulse reactor IBR-2 are
described in Refs. \cite{Borz3,PEPAN}.
 It was decided to carry out a new measurement using higher neutron
flux and with desirably lower background.

 The experiment was performed at the cold neutron prompt gamma activation
analysis facility (PGAA) of the Budapest Neutron Center \cite{PGAA}.
 The beam of cold neutrons was extracted from the cold source of the reactor,
transported through long curved neutron guide, and passed to the target in an
evacuated flight tube.
 The neutron beam cross section on the target was 2$\times$2 cm$^{2}$, the
average thermal equivalent neutron flux at the sample position was about
$5\times 10^{7}$ cm$^{-2}$s$^{-1}$.
 The sample chamber was of $10\times 17\times 22$ cm$^{3}$.
 The flight tube and the target chamber was lined inside with slow
neutron absorber made from $^{6}$Li containing plastic sheet.
 The target materials in our experiments were polyethylene and water.
 The polyethylene target was located between two gamma-ray detectors: a BGO
shielded Compton suppressed n-type coaxial HPGe detector diam.
50$\times$76 mm$^{3}$, and a low energy HPGe detector diam.
35.7$\times$15 mm$^{3}$.
 In the case of water target only the BGO shielded Compton suppressed n-type
coaxial HPGe detector was used.
 The targets were seen by the detectors through the lead collimators.
 The distance between the target center and the coaxial Ge detector was 23.5 cm,
between the target center and the low energy Ge detector -- 16 cm.
 The total measuring time was about 120 hours for each of these targets.
 The peak area of the main H(n, $\gamma$)D transition measured with the coaxial
detector and the polyethylene target was S(2223)=$3.8\times 10^{8}$, the same
with the low energy detector -- $3.7\times 10^{7}$, and $1.45\times 10^{8}$
with the coaxial detector and water target.
 Background spectra were acquired with a graphite target and without any
target as well.

\section{Reduction of data}

 Fig. 1 shows the part of the spectrum of the HPGe coaxial detector in the
energy range 50-2400 keV measured with the polyethylene target.
 The most powerful peaks at the energy of 2223.25 keV (n-p capture $\gamma$-ray),
its single- and double-escape peaks at the energies of 1712 and 1201 keV
respectively, the positron annihilation peak at the energy of 511 keV are
accompanied by many peaks caused by background of gamma-rays generated in the
detector itself, and materials surrounding the target and detectors.

 An energy calibration procedure was used for finding a correspondence between
the peak positions in the spectrum and energies of gamma-rays.
 As a calibration curve we used polynomial of third order consisting of terms
having as parameters the coefficients of the energy calibration curve.
 The least squares method \cite{MINUIT} was used to determine coefficients of
the calibration polynomial.
 22 $\gamma$-peaks arising from neutron capture by hydrogen and nuclei of isotopes
 of Ge, $^{35}$Cl, $^{12}$C, $^{14}$N, $^{27}$Al, $^{207}$Pb present in the
spectrum were used for the calibration in the energy range from 50 keV to 11
MeV.
 Their energy values were taken from the NNDC and IAEA data bases \cite{prompt}.

 The $\gamma$-ray spectrum analysis programs VACTIV \cite{Zlo} and Genie
\cite{Genie} have been used to obtain some of the peak areas in the measured
spectra.
 These programs give contradicting results for the low intensity peaks, the 
latter were analyzed with the MINUIT program \cite{MINUIT}.

  Figures 2-4 show parts of the measured spectra in vicinity of the main
n(p,D)$\gamma$ transition - in the 2100-2210 keV energy range.

 Fig. 2 presents the spectrum measured using the polyethylene target with the
coaxial HPGe detector.
The peak doublet with energies of 2156.4 and 2158.9 keV has special interest.
 Their intensities are $\sim$0.44$\times$10$^{-4}$ and
$\sim$0.8$\times$10$^{-4}$ respectively relative to the main 2223 keV peak.
 A possible interpretation of these peaks could be 2156.3 keV $\gamma$-line in
$^{36}$Cl and 2159.1 keV line in $^{56}$Mn, but their measured intensities are
more than order of magnitude larger than it would follow
(2.3\% and 1.4\% respectively) from the ratio of the detected more intensive
lines in these nuclei after thermal neutron capture in $^{35}$Cl and $^{55}$Mn
\cite{prompt}.
 There was found no other reasonable identification for these $\gamma$-rays:
neither in prompt \cite{prompt} nor in radioactive \cite{rad} gamma-ray
transitions data bases.
 Therefore at least one of these peaks could be considered as a candidate for
second gamma ray in the searched for two-step gamma-ray transition
$^{3}$S$_{1}$(cont.)$\rightarrow^{1}$S$_{0}$(metastable)$\rightarrow^{3}$
S$_{1}$(ground state).

 Fig. 3  shows the same part of the spectrum of the low energy planar HPGe
detector measured with the polyethylene target.
 This spectrum contains a number of interpreted weak peaks (see below) but none
of two peaks (2156.4 and 2158.9 keV) was found in the spectrum of the low
energy detector having (due to an order of magnitude lower background)
sensitivity to weak peaks in this energy range not much worse than the coaxial
detector.
 Therefore we consider these peaks as due to $\gamma$-rays coming from the
surrounding of the coaxial detector or originating in the detector itself.

 The main difficulty in the interpretation of the experimental spectra was
caused by the background of gamma-rays generated in detectors and material
surrounding the target and the detectors.
 Numerous peaks from neutron capture by isotopes of Ge, Cl, Fe, and by nuclei
$^{27}$Al, $^{12}$C, $^{14}$N etc have been observed in all measured spectra.
 We detected $\sim$250 peaks in the total spectrum up to 11 MeV.

 From the area of the photo-peak with E=2223 keV: S=3.8$\times$10$^{8}$ measured
with the polyethylene target, known photo-peak efficiency 1.9$\times$10$^{-4}$
at this energy and distance from the target to the detector, and the ratio of
the cold neutron-proton scattering to capture cross sections ($\sim$50 b/0.8
b$\approx$60) we estimate the total number of cold neutrons scattered by the
polyethylene target as $\sim 2\times 10^{14}$.
 From the areas of numerous $\gamma$-peaks identified as arising from neutron 
capture by Ge isotopes in the coaxial HPGe detector and the BGO anti-Compton 
shield and assuming that these peaks were due to capture of cold neutrons 
scattered by the polyethylene target we estimated the fluence of cold neutrons 
at the coaxial detector to be about $\sim$10$^{5}$ cm$^{-2}$.

 In addition, as is known \cite{nLi6} $^{6}$Li-containing materials produce
fast neutrons at energies about 16 MeV after neutron capture by $^{6}$Li
through the reactions:
\begin{equation}
n(^{6}Li,\alpha)t\quad (E_{t}=2.73 MeV);  \qquad t(^{6}Li,^{8}Be)n, \quad
Q=16.02 MeV; \quad t(^{6}Li,2\alpha)n, \quad Q=16.15 MeV
\end{equation}
with probability $\sim$10$^{-4}$ per one thermal neutron captured by
$^{6}$Li nuclei.
 As practically all cold neutrons scattered by targets are captured in the
$^{6}$Li shielding we estimate the total fluence of fast neutrons irradiating
the HPGe detector as 10$^{14}\times$10$^{-4}$/4/3.14/23.5$^{2}$=
1.5$\times$10$^{6}$/cm$^{2}$.
 Fast neutrons can produce numerous reactions of the type: (n,n$^{'}$),
(n,p), (n,$\alpha)$, (n,2n), etc in HPGe detector and surrounding materials.
 Nuclear data concerning gamma-rays from nuclei produced in result of these
reactions are scares.
 Therefore it is not surprising that we found more than two dozens of
unidentified peaks in our gamma spectra.

 For purposes of more detailed analysis the spectra and obtaining constraints
on the searched for two-photon transition the energy region of interest
2100-2210 keV has been divided into several parts.
 For example the spectrum in this energy range measured with the coaxial HPGe
detector and polyethylene target (Fig. 2) was divided into five parts:
2100-2130 keV, 2130-2152 keV, 2152-2164 keV, 2164-2180 keV and 2180-2210 keV.
 In each of these regions spectrum was described as a sum of polynomial
functions up to third order:
\begin{equation}
N_{1}(E)=a_{0}+a_{1}E+a_{2}E^{2}+a_{3}E^{3}
\end{equation}
and Gaussian with $\sigma$ corresponding to the HWHM $\Delta$=1.3 keV of the
main peak with E=2223 keV:
\begin{equation}
N_{2}(E)=b_{1}exp(-((E-b_{2})/\sigma)^{2}).
\end{equation}

 The least-squares MINUIT program \cite{MINUIT} was used to determine
constraints on the magnitude of possible gamma-ray peak $b_{1}$.

 The spectrum of the low energy detector (Fig. 3) contains peaks with energy
2108.2 keV identified as the gamma-transition in $^{28}$Al after neutron
capture by $^{27}$Al, sum of 2111.9 keV ($^{74}$Ge) and 2113.4 keV ($^{57}$Fe),
2127.1 keV ($^{74}$Ge), 2138.8 keV ($^{28}$Al), and 2170.7 keV ($^{28}$Al).
 The peak 2163.5 keV did not find interpretation, but these peaks are hardly 
visible in the spectrum of the coaxial detector except for the weak peak at 
2138.8 keV ($^{28}$Al).
 Thus the use of two detectors in the experiment permitted to exclude two-gamma
cascade from np-capture.

 Fig. 4 shows the spectrum in the energy range 2100-2210 keV measured with
the water target.
 The peak 2184 keV is identified as due to neutron capture by $^{16}$O in water,
the weak peak at 2108 keV from $^{28}$Al nucleus.
 Their areas are in good agreement with the calculated ones, the first one -
taking into account the ratio of the neutron capture cross sections by $^{16}$O
and by protons in the water target and the probability 82\% of the
gamma-transition in $^{17}$O, the second from comparing its area with the areas
of more intensive peaks from $^{28}$Al \cite{prompt}.

 Using the same procedures for all spectra shown in Figs. 2-4 we obtained
the final constraints R for the ratio of the magnitude of the searched for
second peak corresponding to the two-step gamma-ray transition
$^{3}$S$_{1}$(continuum)$\rightarrow^{1}$S$_{0}$ (metastable)
$\rightarrow ^{3}S_{1}$ (ground state) to the main transition
$^{1}S_{0}\rightarrow^{3}S_{1}$ with the energy 2223 keV: $R<6\times
10^{-6}$, and the cross section of the $np$ capture followed by such a
transition $\sigma<2\mu$b (two standard deviations).

\section{Conclusion}
 Our result implies that there is no evidence for the two-photon transition in
the $np$ capture with one of gamma-rays in the region 2100-2210 keV.
 The branching ratio is $R<6\times 10^{-6}$ or cross section $\sigma<2\mu$b
(two standard deviations).
 This value is more than an order of magnitude less than prediction in
\cite{Hack}.
 Although this limit rejects the prediction of \cite{Hack} it still leaves room
for further investigations with desirably lower background.

\newpage
\begin{figure}
\begin{center}
\resizebox{13cm}{13cm}{\includegraphics[width=\columnwidth]{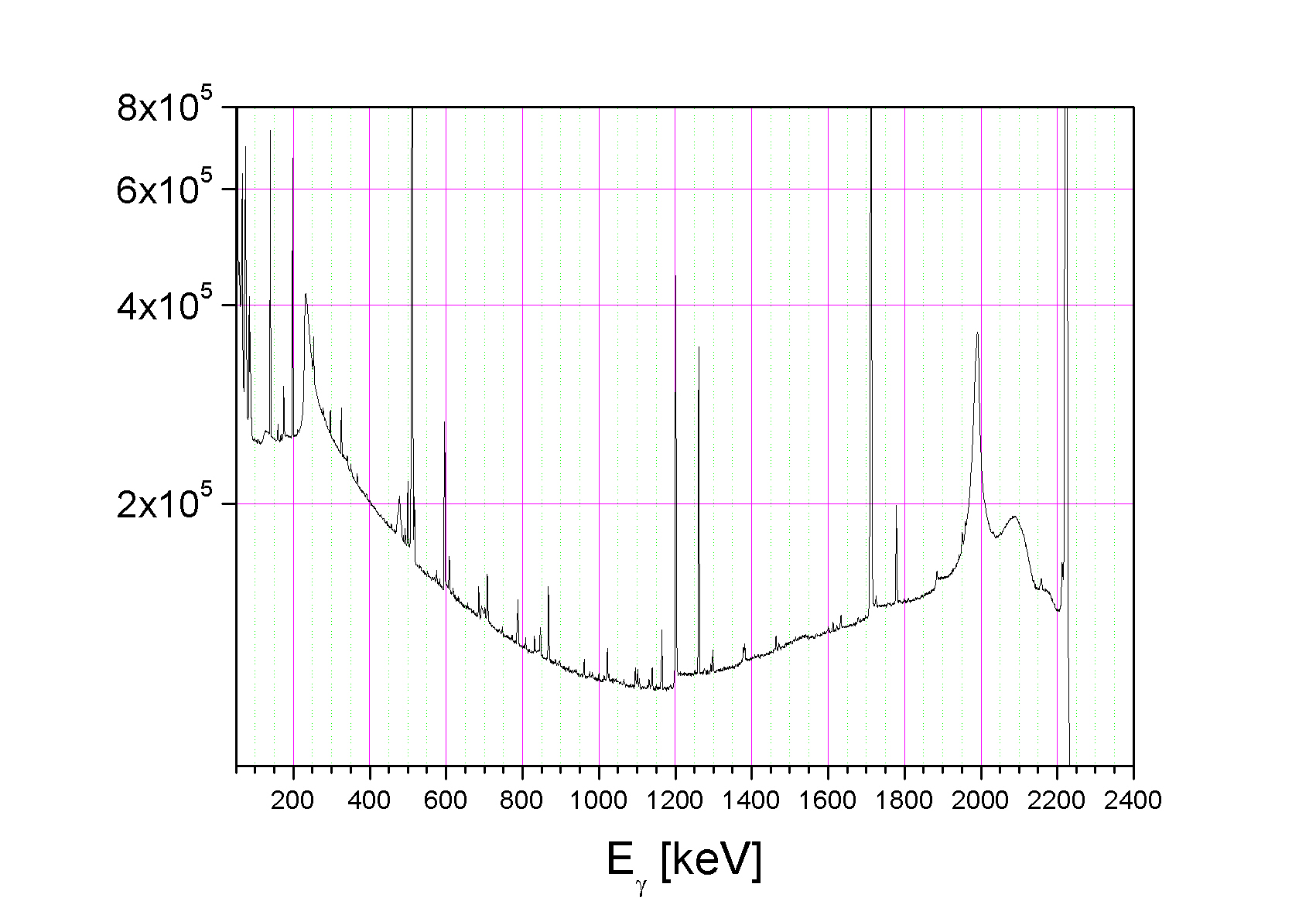}}
\end{center}
\caption{Part of the gamma ray spectrum (50-2400 keV) of the coaxial HPGe
detector measured with the polyethylene target.}
\end{figure}

\newpage
\begin{figure}
\begin{center}
\resizebox{13cm}{13cm}{\includegraphics[width=\columnwidth]{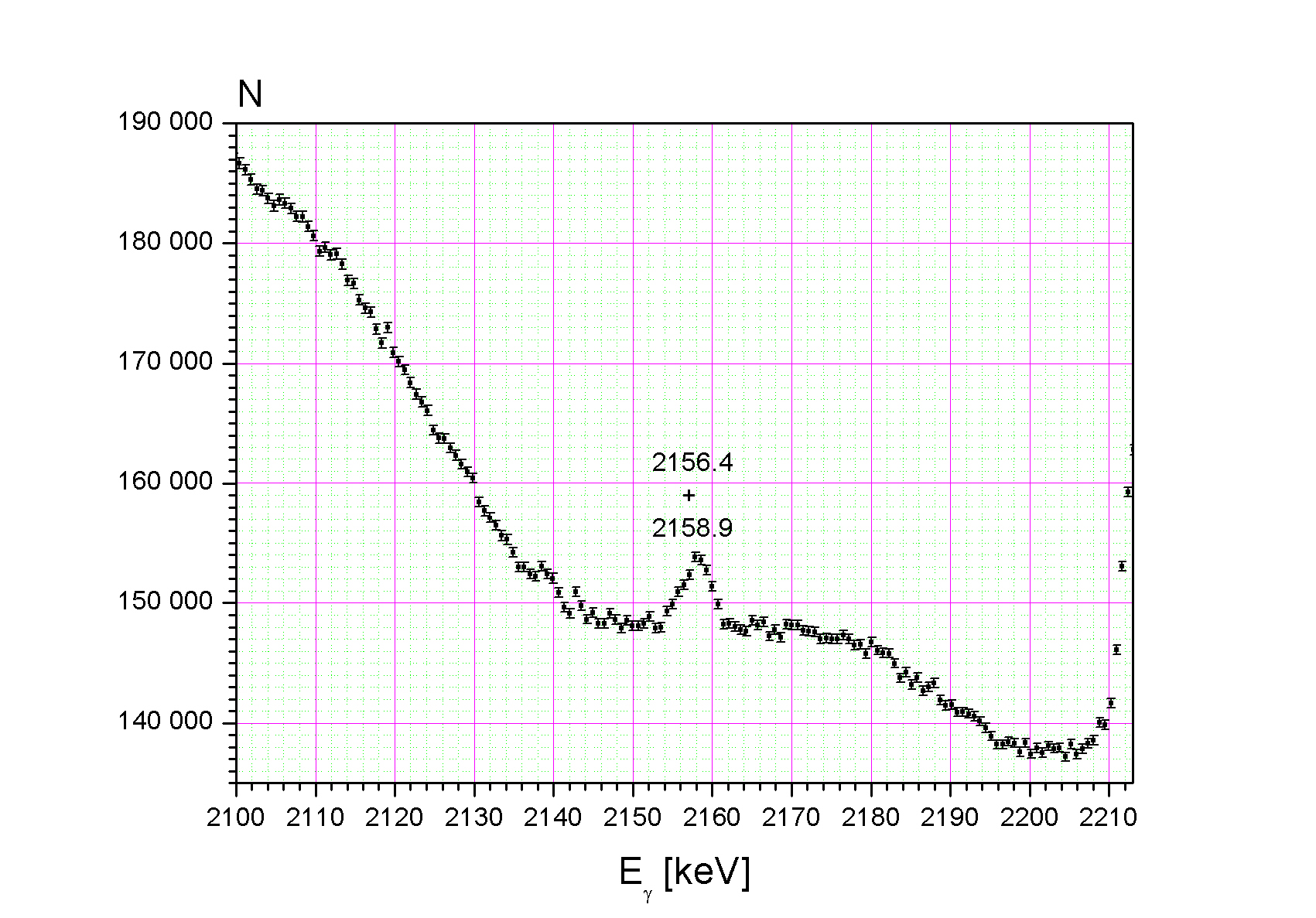}}
\end{center}
\caption{Part of the gamma ray spectrum (2100-2210 keV) of the coaxial HPGe
detector measured with the polyethylene target.}
\end{figure}

\newpage
\begin{figure}
\begin{center}
\resizebox{13cm}{13cm}{\includegraphics[width=\columnwidth]{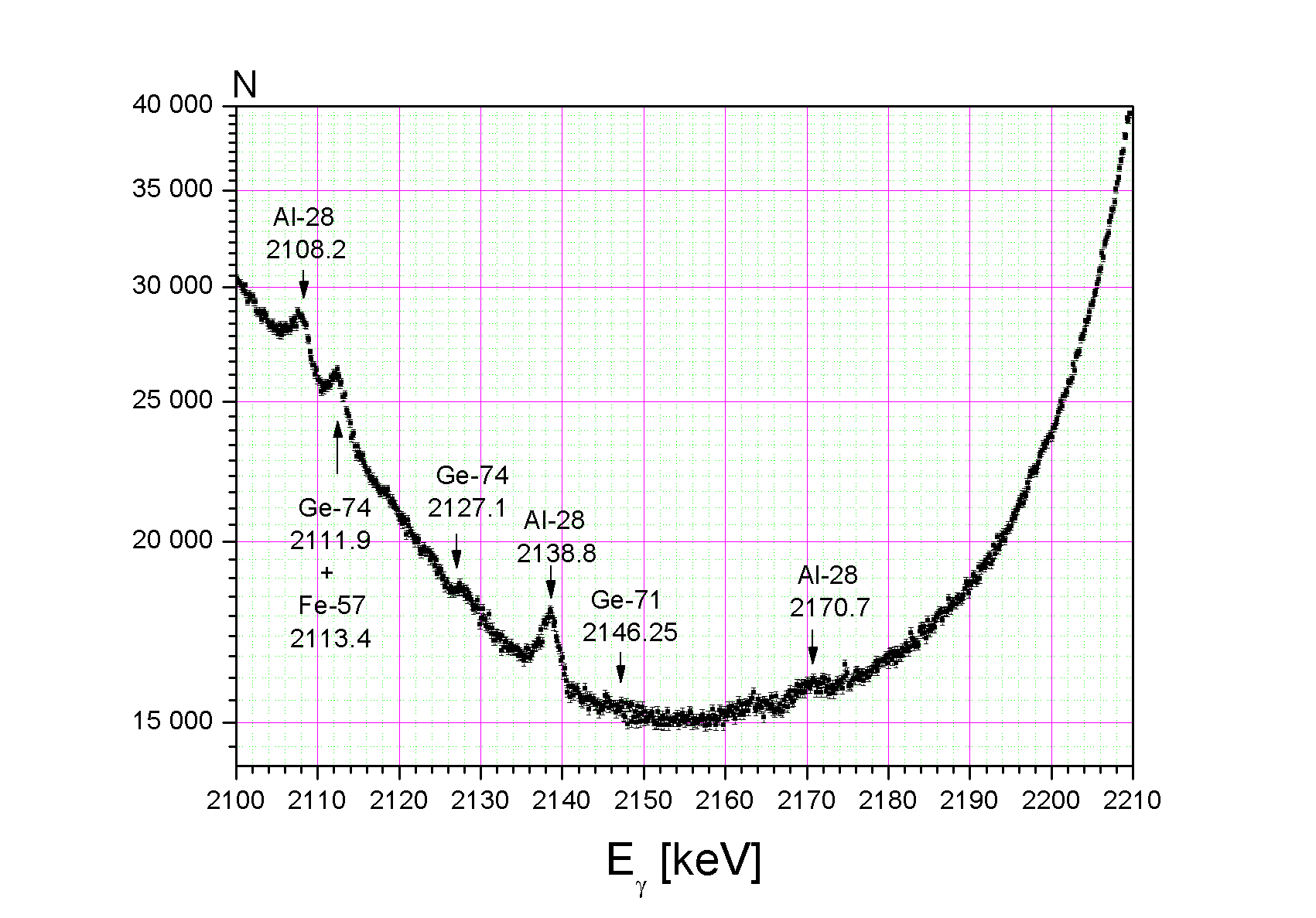}}
\end{center}
\caption{Part of the gamma ray spectrum (2100- 2210 keV) of the low energy HPGe
detector measured with the polyethylene target.}
\end{figure}

\newpage
\begin{figure}
\begin{center}
\resizebox{13cm}{13cm}{\includegraphics[width=\columnwidth]{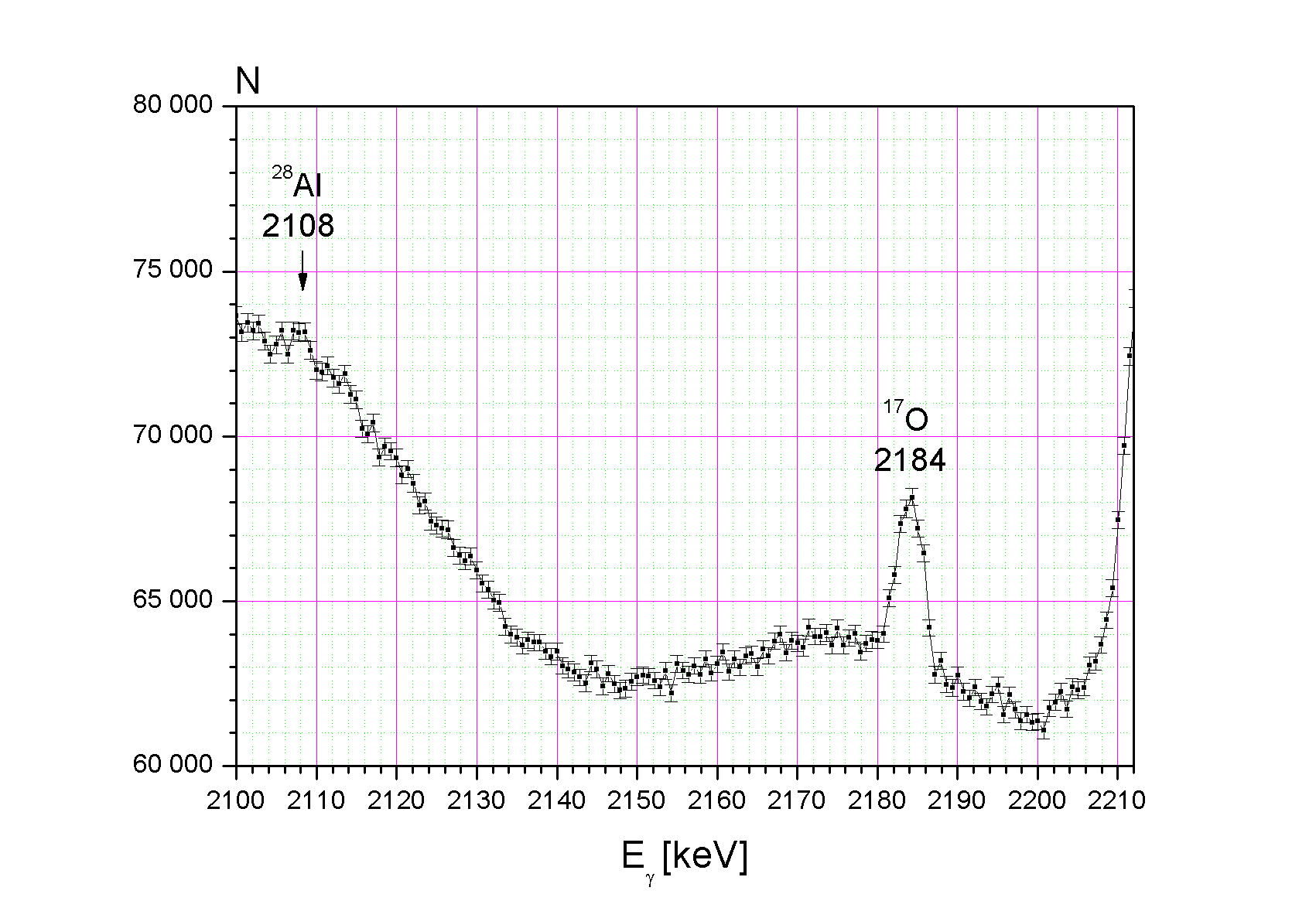}}
\end{center}
\caption{Part of the gamma ray spectrum (2100-2210 keV) of the coaxial HPGe
detector measured with the water target.}
\end{figure}

\end{document}